\RequirePackage{fix-cm}
\documentclass[smallextended]{svjour3}       
\smartqed  

\newcommand*{\affaddr}[1]{#1} 
\newcommand*{\affmark}[1][*]{\textsuperscript{#1}}

\usepackage{graphicx}
\graphicspath{{./p_images_final/}}
\usepackage[percent]{overpic}

\usepackage{lineno}
\usepackage{amssymb}
\usepackage{amsmath}
\usepackage{placeins}
\usepackage{longtable}
\usepackage{setspace}
\usepackage{tikz}
\usepackage{lscape}
\usepackage{array}
\usepackage[square,numbers]{natbib}
\newcolumntype{M}{>{\centering\arraybackslash}m{2cm}}
\usepackage{gensymb}
\usepackage{bm}

\begin{document}

\title{On assuring the accurate alignment of laser sheets for planar and stereoscopic PIV}

\titlerunning{Laser sheet alignment for planar and stereoscopic PIV}        

\author{Muhammad Shehzad\affmark[1]  \and
        Sean Lawrence\affmark[1]  \and
        Callum Atkinson\affmark[1]  \and
        Julio Soria \affmark[1] 
}

\authorrunning{Shehzad et al. 2021} 

\institute{Muhammad Shehzad \at
\email{Muhammad.shehzad@monash.edu} \and
\affaddr{\affmark[1]Laboratory for Turbulence Research in Aerospace \& Combustion (LTRAC) Department of Mechanical and Aerospace Engineering, Monash University, Victoria 3800, Australia}}

\date{Received: date / Accepted: date}

\maketitle

\thispagestyle{plain}

\begin{abstract} 
Several calibration techniques have been proposed in the literature for the calibration of two-component two-dimensional (2C-2D) particle image velocimetry (PIV) and three-component two-dimensional (3C-2D) stereoscopic PIV (SPIV) systems. These techniques generally involve the use of a calibration target that is assumed to be at the exact centre of the laser sheet within the field of view (FOV), which in practice is very difficult to achieve. In 3C-2D SPIV, several methods offer different correction schemes based on the computation of a disparity map, which are aimed at correcting errors produced due to this misalignment. These techniques adjust the calibration of individual cameras to reduce the disparity error, but in doing so can create unintended errors in the measurement position and/or the velocity measurements, such as introducing a bias in the measured three-component (3-C) displacements. This paper introduces a novel method to ensure accurate alignment of the laser sheet with the calibration target so that the uncertainty in displacement measurements is less than or equal to the uncertainty inherent to the PIV and hence, no correction scheme is required. The proposed method has been validated with a simple experiment in which true displacements are given to a particle container (illuminated by an aligned laser sheet) and the measured 3C displacements are compared with the given true displacements. An uncertainty of less than $7.6\,\mu m$ (equivalent to $0.114\,px$) in the measured 3C displacements demonstrates the effectiveness of the new alignment method and eliminates the need for any ad hoc post-correction scheme.
\end{abstract}



\section{Introduction}
\label{sec:introduction}

Calibration is important for planar (2C-2D) and stereoscopic (3C-2D) PIV to determine magnification, image distortions, camera orientation and position. Sometime this information is explicitly determined ({\em e.g.} focal length, camera positions in the camera pinhole model \cite{tsai1987versatile}) or may be combined and embedded with a mapping function (polynomial \cite{soloff1997distortion, Oord1997SPIVsystem}, rational functions \cite{willert1997stereoscopic}). During the calibration, images of a calibration target are obtained while it is placed in the position of the laser sheet within the field of view (FOV). The calibration target could have a chequerboard print or a grid of regularly spaced markers. A mapping function is then used to map the object (global) space to the image space. We define $(X,Y,Z)$ as the Cartesian coordinate system of the object space and $(x,y,z)$ of the image space. 

\begin{figure}
\begin{center}
\begin{tabular}{cc}
   (a)  & (b) \\
\includegraphics[width=0.37\textwidth]{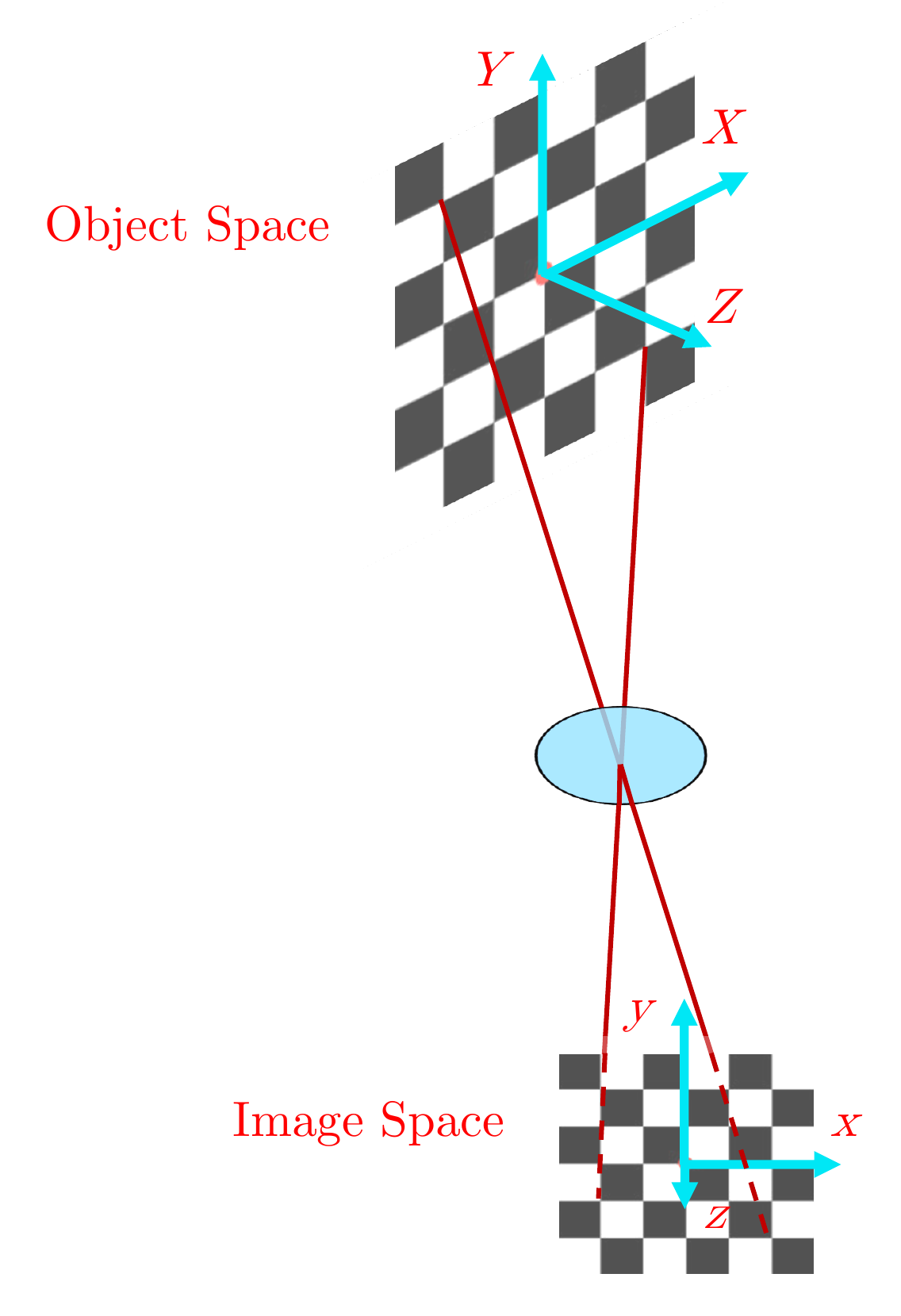}
     &     \includegraphics[width=0.64\textwidth]{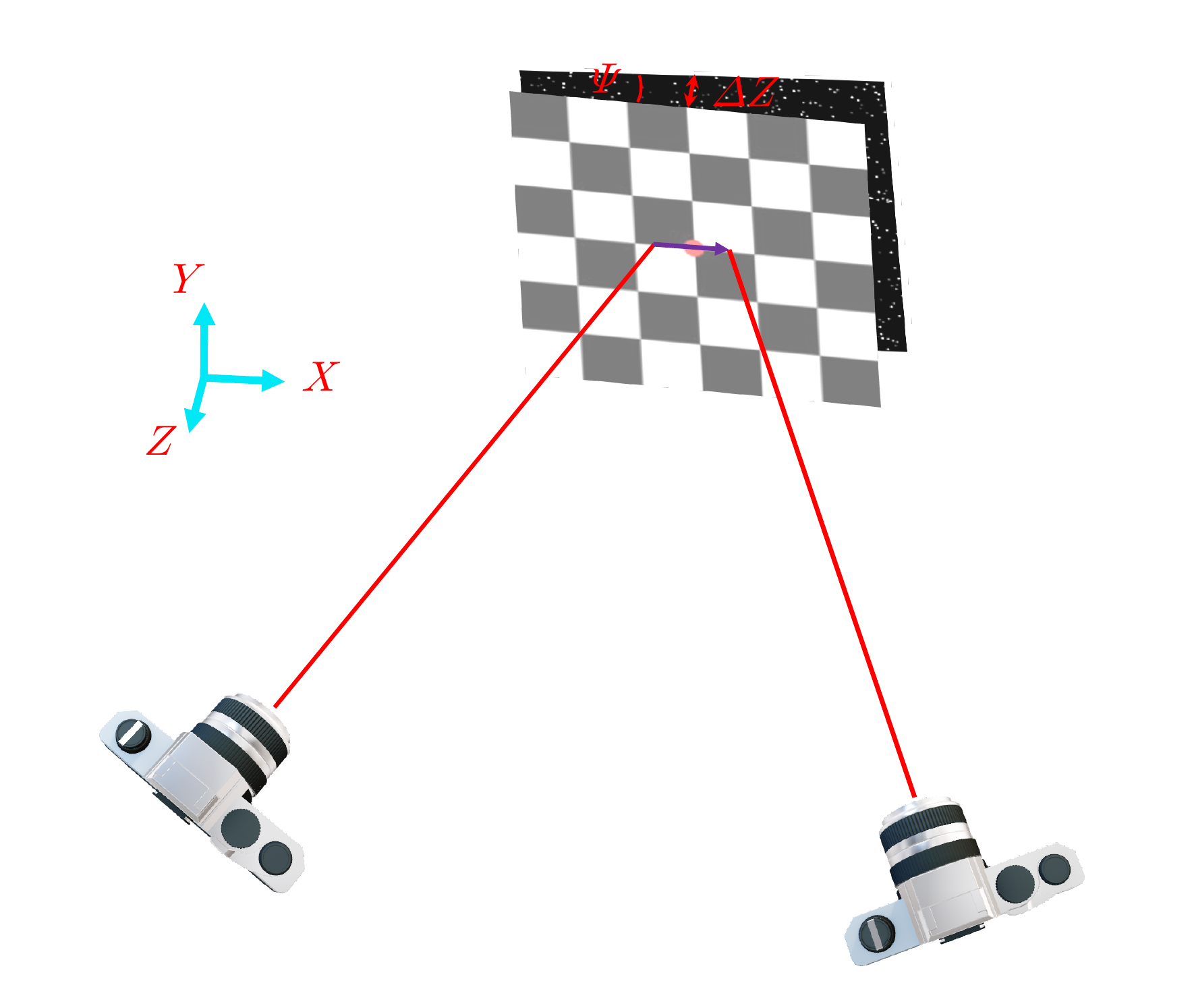} \\
\end{tabular}
\end{center}
\caption{(a) A schematic of 3C-2D SPIV system showing the object and image spaces (b) An illustration of the misalignment in the form of rotation or offset between the laser sheet and the calibration target.}
\label{fig:spiv_coordinates_and_misalignment}    
\end{figure}

In 2C-2D PIV, a polynomial mapping function requires only one image of a planar calibration target at a particular $Z$ location, to map $(X,Y)$ to $(x,y)$. This is referred to as 2D calibration. In 3C-2D SPIV (see illustration in figure \ref{fig:spiv_coordinates_and_misalignment}(a)), the calibration needs the global $Z$ dependency to determine the out-of-plane component of velocity. It, therefore, needs the calibration target to be imaged at several (minimum two) known $Z$ locations which are parallel to each other. Alternatively, a 3D calibration target can be used which eliminates the need for the $Z$ translations \cite{wieneke2005stereo}. A comparison between the performance of 2D and 3D calibration targets in 3D calibration has been presented in \citet{scarano2004comparison} and the authors suggest that the 2D calibration targets lead to more accurate results because of the higher spatial resolution they provide. This is because a 3D grid needs to show markers in separate planes, which limits the marker density in a single plane. This calibration with global $Z$ dependency is known as 3D calibration \cite{prasad2000stereoscopic} which involves the use of a mapping function (pinhole model, polynomial or rational function) that maps a position in the global coordinate system $(X,Y,Z)$ to a location in the image plane $(x,y)$ of a given camera. The reconstruction of the 3C-2D vector field from the 2C-2D vector fields obtained with the PIV analysis of the dewarped image pairs of the individual cameras is performed following the methods of \citet{soloff1997distortion} or \citet{willert1997stereoscopic}. The global $Z$ dependency of the mapping function is used during the reconstruction. 

If the calibration target is not placed precisely at the position of the measurement plane in the laser sheet, this is referred to as \textit{misalignment}. The misalignment leads to errors that are inherent to the aforementioned calibration methods \cite{prasad2000stereoscopic}. These errors are classified as \textit{perspective error} in 2C-2D planar PIV and \textit{position error} and \textit{3C-reconstruction error} in 3C-2D SPIV \cite{giordano2009spatial}. In 3C-2D SPIV, the misalignment leads to two different positions of the same point in the object space viewed by the two cameras and reconstruction of 3C displacement vectors from 2C vectors is made relative to those positions which, therefore, causes the position error. The 3C-reconstruction error arises during the 3C-reconstruction of 
\citet{soloff1997distortion} due to computation of local velocity gradients at erroneous positions and \citet{willert1997stereoscopic} because of an error in the calculation of the viewing angle due to incorrect positions. 

Misalignment between the measurement plane and the calibration target can take the form of a rotation about $X$ or $Y$ axis and/or an offset in $Z$ direction as illustrated in figure \ref{fig:spiv_coordinates_and_misalignment} (b) where $\Psi$ and $\Delta Z$ represent the angle of rotation and the $Z$ offset, respectively. A small rotation misalignment of $0.6 \degree$ results in disparity vectors as large as $1 \, mm$ near the edges of the FOV \cite{willert1997stereoscopic}. The disparity vectors or the disparity map is obtained from the cross-correlation between the particles images of camera 1 and camera 2, dewarped with their respective mapping functions, to obtain an estimation of nature and the magnitude of the misalignment. Similar results have been reported by \citet{coudert2001back,van2004measurement} and \citet{wieneke2004application} among others. In the case of a $Z$ offset, for example, an offset of $20 px$ and a velocity gradient of $5\%$ would generate an error of $5\%$ in the 3C displacement measurement \cite{wieneke2005stereo}.

A target free calibration method proposed by \citet{fouras2007simple} and further detailed in \citet{fouras2008target} suggests the elimination of the calibration target and use of a third paraxial camera instead to compute mapping functions of each of the two stereo cameras from their particles images cross-correlated with the particle images of the paraxial camera, all recorded at the same time. While this method seems attractive when placement of a calibration grid in the measurement region is not feasible, it, however, needs an additional camera of the same spatial resolution as the other two and hence adds to the cost of the experimental setup. Moreover, the computation of mapping functions from the cross-correlation of the particle images from the two cameras with those from the paraxial camera is prone to errors produced because of the possible erroneous point-correspondences, as the particle size is much smaller than that in a usual calibration grid.

To the best of the authors' knowledge, no method in the literature proposes any technique to ensure and evaluate alignment between a calibration target and the laser sheet. This paper introduces a novel method to achieve alignment between the calibration target and the laser sheet and eliminate a major source of errors: the misalignment. The new method suggests a simple, low-cost and easy-to-use technique to achieve alignment and thereby reduce both the error introduced by misalignment and any subsequent bias errors that might be introduced by the use of a post-measurement correction scheme. This method is introduced in section \ref{sec:new_alignment_method}. The details of the experimental validation of the new alignment method are presented in section \ref{sec:validation}. Section \ref{sec:summary} summarizes the paper along with concluding remarks.

\section{A new method of alignment}
\label{sec:new_alignment_method}

A major challenge in achieving correct calibration and thereby obtaining sufficiently accurate 2C-2D or 3C-2D SPIV measurements is to align the laser sheet with the calibration target. If we can achieve alignment up to a satisfactory level, the uncertainty in the 3C displacement measurements will reduce significantly. This paper proposes a novel method to achieve practical alignment. The key to this method is the provision of a mechanism through which the position of the laser sheet can be adjusted to ensure that its mid-plane is coincident with the calibration target plane. For this purpose, it is desirable to have a calibration target that has the capability of objectively determining the parallelism of the laser sheet relative to the target. For this purpose, a hollow rectangular calibration target frame is constructed, which lets the light sheet pass through one side and enables its detection on the opposite side. To remove the need for manual pinhole type alignment, two or more photodiodes are used to detect the intensity of light that passes through the incident side and arrive at the opposite side of the frame. This frame is referred to as a calibration target mount and is the main component of the alignment setup.

\begin{figure}[tbph]
\begin{center}
\begin{tabular}{cc}
(a) & (b) \\
\includegraphics[trim={0cm 0cm 0cm 0cm},clip,width=0.45\textwidth]{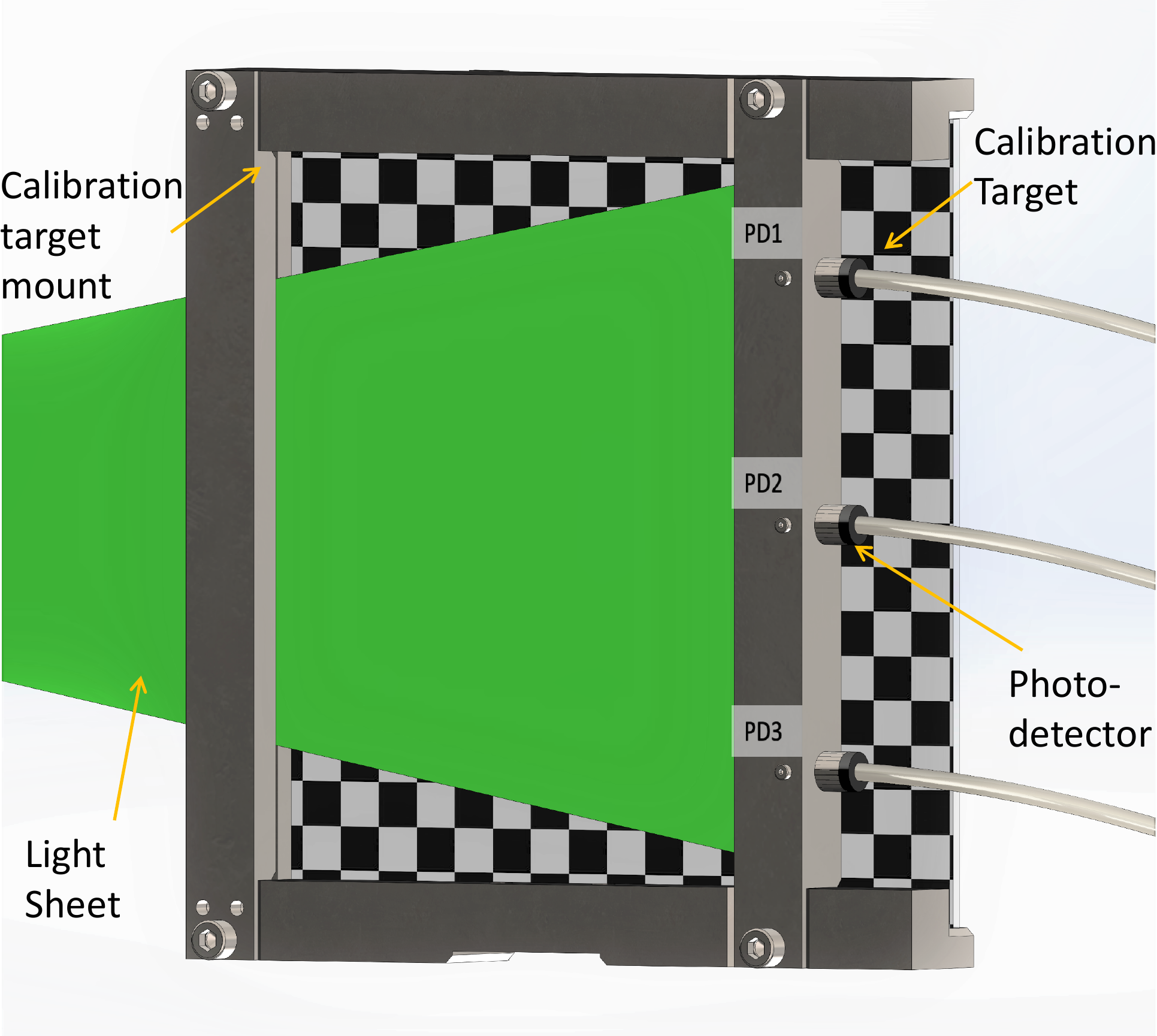} & \includegraphics[trim={0cm 0cm 0cm 0cm},clip, width=0.45\textwidth]{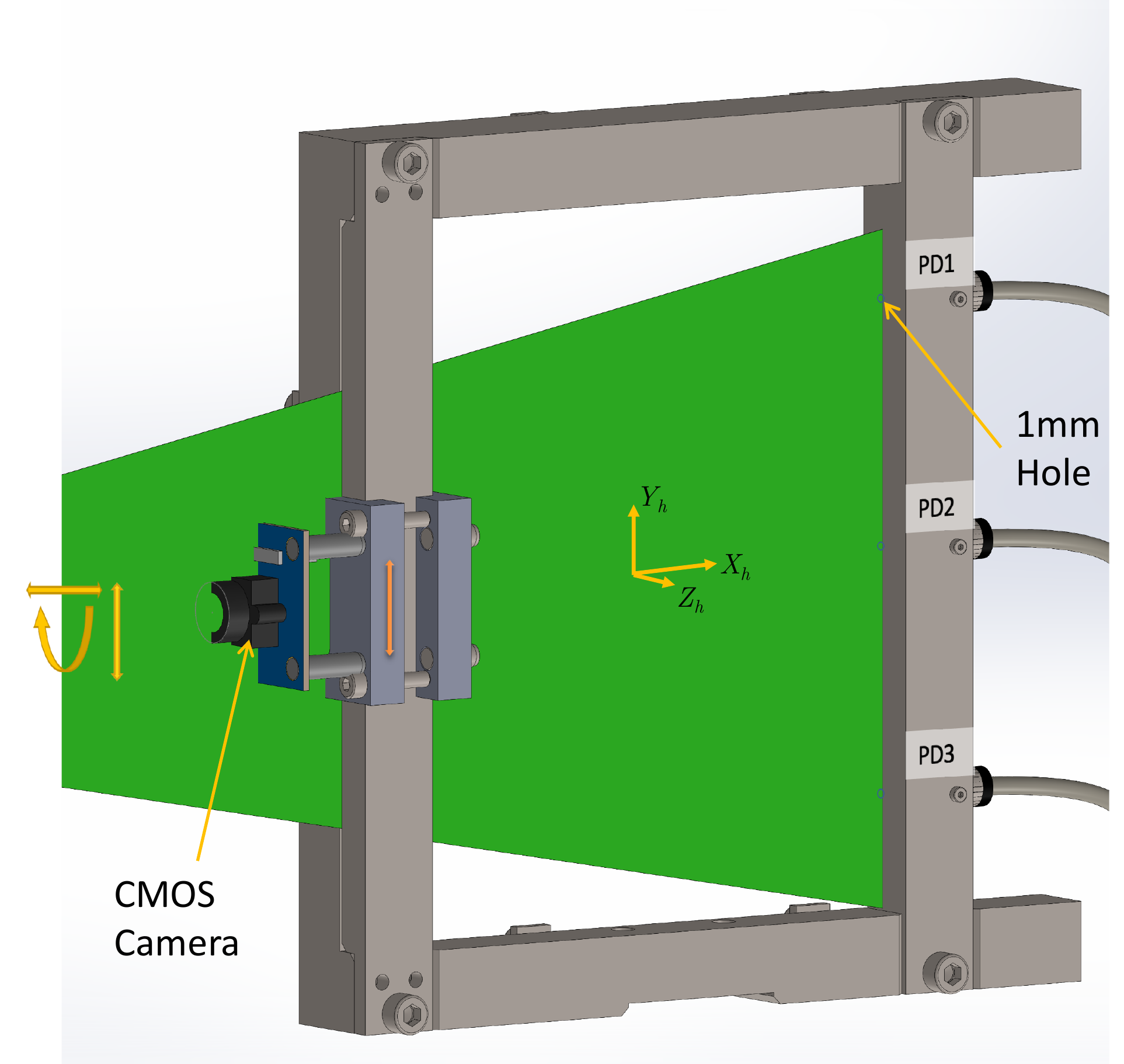} \\
\label{fig:total_alignment_setup}
\end{tabular}
\end{center}
\caption{(a) Parametric view of the laser sheet alignment setup (b) Parametric view of the alignment verification setup}
\end{figure}

For the implementation of this principle, an alignment setup as shown in the schematic in figure \ref{fig:total_alignment_setup}(a) has been built. A calibration target mount has been designed and manufactured at Laboratory for Turbulence Research in Aerospace and Combustion (LTRAC). The inner open face of the target mount has dimensions of $181 \times 141 \, mm^2$. The mount holds the calibration target fixed in position using six fitting screws. It has a 1 $mm$ slit on the front side to let a laser sheet pass through it and fall on three 1 $mm$ diameter vertically equidistant holes at $Y_h=[+62,0,-62] \, mm$ on the inner face of the rear side. (These holes are referred to as the H\{1,2,3\} and $(X_h, Y_h)$ is the Cartesian coordinate system of the calibration target with its origin at the middle of the open face of the mount. The offset of the slit from the front face of the calibration target when it is fixed in the target mount is $8.5mm$. After aligning the laser sheet with the target mount, the mount needs to be translated by this offset to bring the calibration target to the position of the laser sheet before acquiring the calibration images. The target mount has three cylindrical housings on the outer face of the rear side (behind H\{1,2,3\}) to hold three photo-detectors (photo-diodes, referred to as PD\{1,2,3\}) encapsulated in hollow cylinders of 10 $mm$ outer diameter. These PDs, connect to a multi-channel oscilloscope and detect the intensity of laser light falling on them through the 1 $mm$ holes of the target mount, which is proportional to the laser sheet alignment. The intensity measurement is indirect, in the form of a voltage. Hence, the term `Intensity' is used in the text and figures, as a proxy for the voltage measured. 

The process for aligning the laser sheet with the target mount is as follows:
The laser beam, before the sheet formation, is reflected by two mirrors each on a mount with adjustment screws to shift the laser beam propagation axis horizontally or vertically along the directional arrows highlighted in yellow in figure \ref{fig:total_alignment_setup}(b). A rotation stage, on which the sheet forming cylindrical concave lens is mounted, allows the sheet to be rotated with an accuracy of $0.014 \degree$. 
To achieve alignment, the beam propagation axis is aligned with the H2 i.e. at $Y_h=0$  using adjustments screws on laser mirror mounts. A sheet is created by inserting the cylindrical lens into the beam path, rotating the lens within a rotational stage,  to align the sheet to the top and bottom holes (H1,3). 

\begin{figure}
  \centering
      \includegraphics[width=0.65\textwidth]{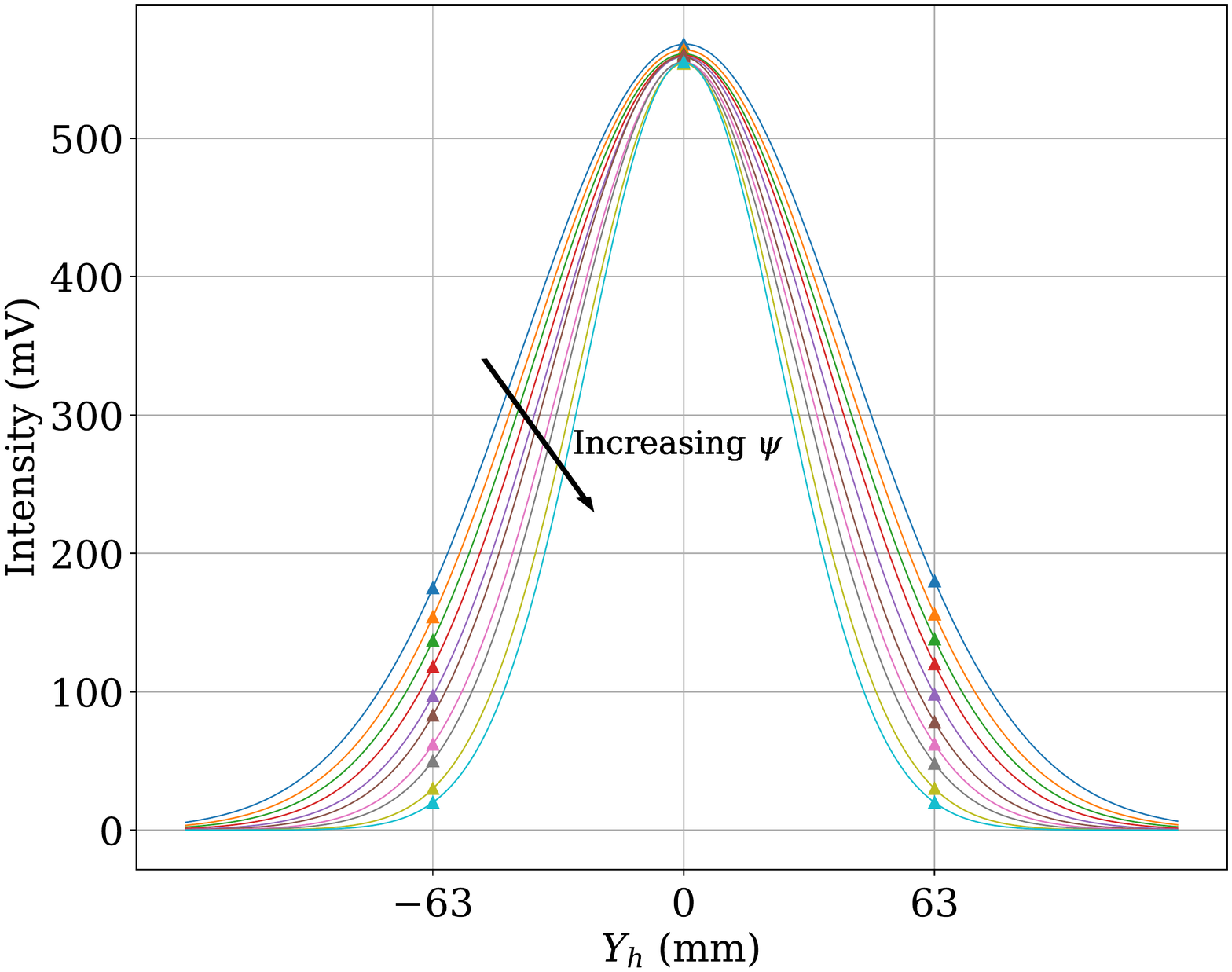}
     \caption{Intensities measured at PD\{1,2,3\} for rotation of laser sheet by increments of $0.1 \degree$. The angle of rotation is increasing in the direction of the black arrow.}
\label{fig:sheet rotation}    
\end{figure}

\begin{figure}
  \centering
      \includegraphics[width=0.7\textwidth]{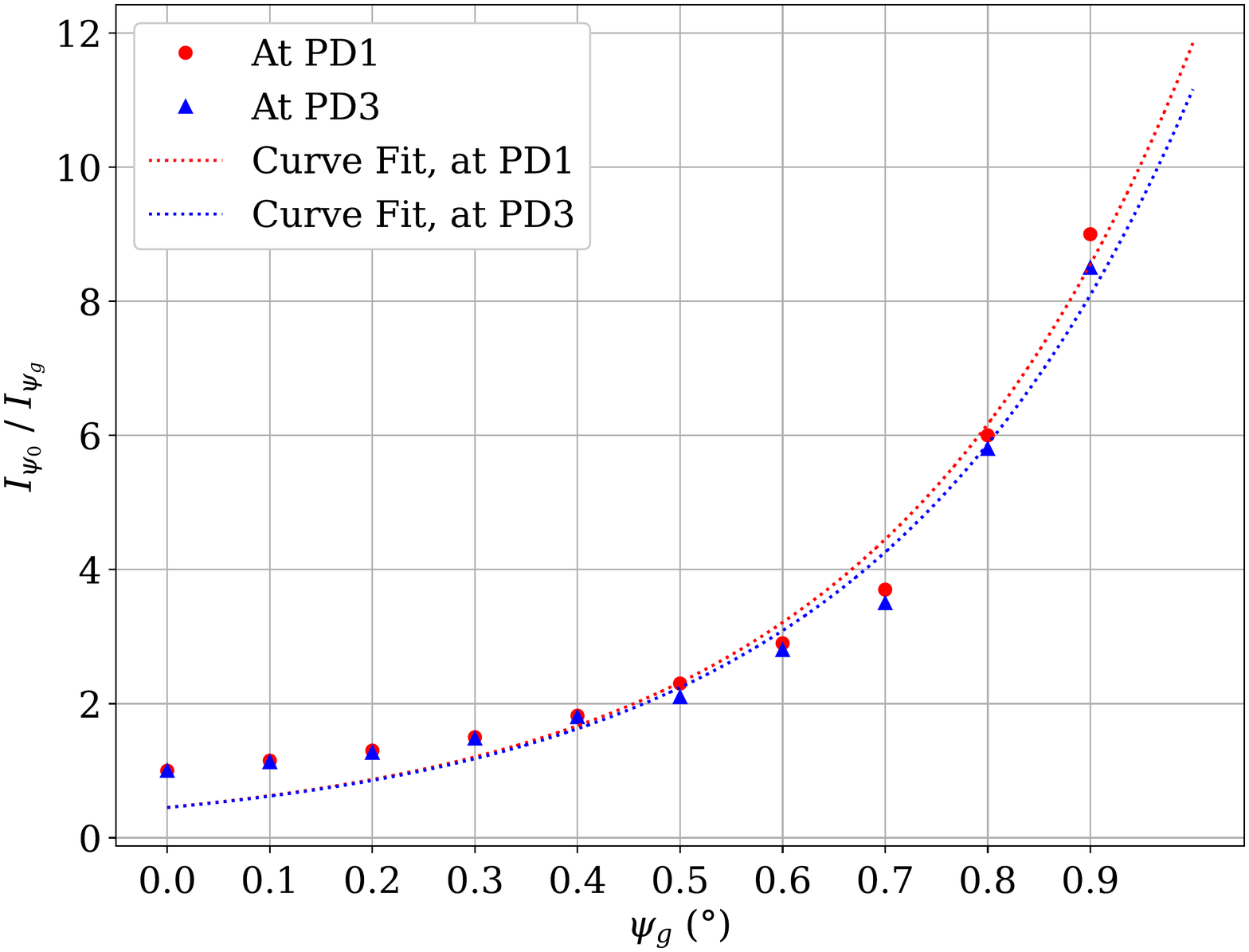}
     \caption{Ratio of the intensities measured at PD1 and PD3 at different sheet rotation angles to their intensities at the reference position  $\Psi_0$.}
\label{fig:detection_ratio}    
\end{figure}

The sheet, after the rotation and the lateral adjustments, is considered aligned with the target mount when maximum values of intensity are measured at all PDs and the intensity values at PD1 (at $Y_h = +62 \, mm$) and PD3 (at $Y_h = -62 \, mm$) are sufficiently close to each other (within $\pm10 mV$ as this is the maximum deviation among the sensitivity values of all PDs at the used laser energy). This position of the laser sheet is taken as a reference position, $\Psi_0$, where $\Psi$ is the angle of the laser sheet rotation with respect to the calibration target mount in $x$ axis. The thickness of the laser sheet is maintained at about 1 $mm$ (using the telescope principle of optical lenses) before it enters the slit in the front side of the target mount. Rotating the laser sheet slightly about the expansion axis by increments of $0.1 \degree$ with respect to the reference position, while keeping the pivot point of the sheet rotation at the H2, will decrease the intensity values measured at PD1 and PD3 while at PD2, the intensity will remain constant. If the rotation is greater than $0.92 \degree$, almost no intensity is expected to be detected at PD1 and PD3 because the laser sheet thickness is only $1 \, mm$ following the $1 \, mm$ slit on the front face. This is evident in figure \ref{fig:sheet rotation} where the sheet is rotated from the  $\Psi_g = 0.1\degree$ to $0.9 \degree$ and the intensity values at PD1 and PD3 drop to near-zero values. The ratio of the intensities at PD1 and PD3 after every rotation relative to their intensities at $\Psi_0$ increases with the increasing sheet rotation angle as shown in figure \ref{fig:detection_ratio}, until it becomes undefined beyond $\Psi_g =0.92 \degree$. This increase is almost exponential as shown by the exponential curve fits included in the figure. Moving the laser sheet slightly along the $Z$ axis will decrease the intensities of all PDs while moving along a positive $Y$ axis would decrease intensities at PD2 and PD3 and increase at PD1. This sensitivity method assists in choosing the optimal laser sheet position relative to the calibration target and is therefore considered as the position of the light sheet where it is parallel to the calibration target.  Keeping the laser sheet fixed now and traversing the target mount by $Z=8.5\pm0.01 \, mm$ using a micrometre translation stage would bring it in alignment with the middle of the light sheet. 

\begin{figure}
  \centering
	\includegraphics[ width=0.65\textwidth]{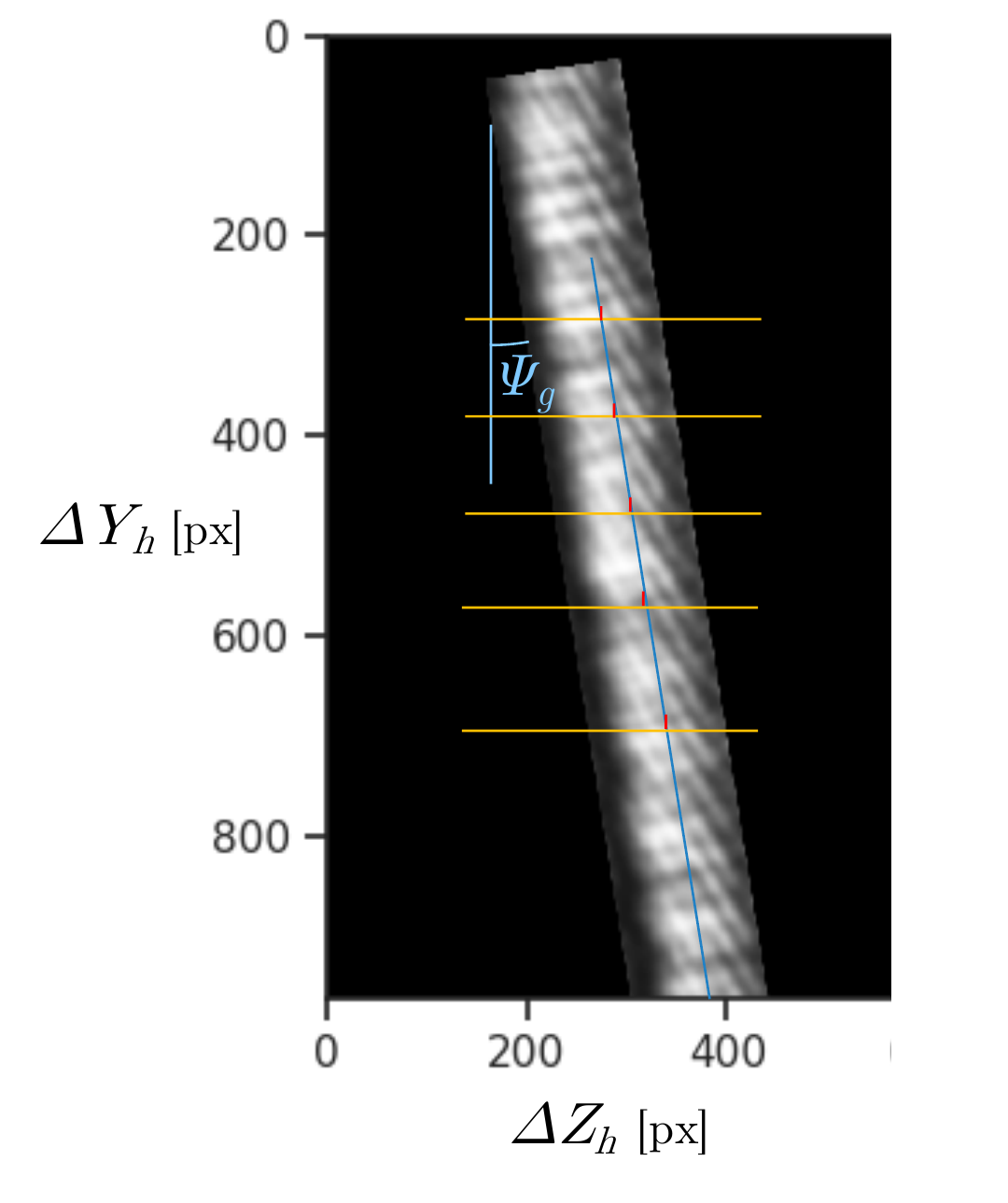}
\caption{Angle of rotation given to the laser sheet and location of the SIP profiles (yellow lines).}  
\label{fig:angle_given_and_SIP_locations}
\end{figure}

\begin{figure}
  \centering
\includegraphics[ width=0.65\textwidth]{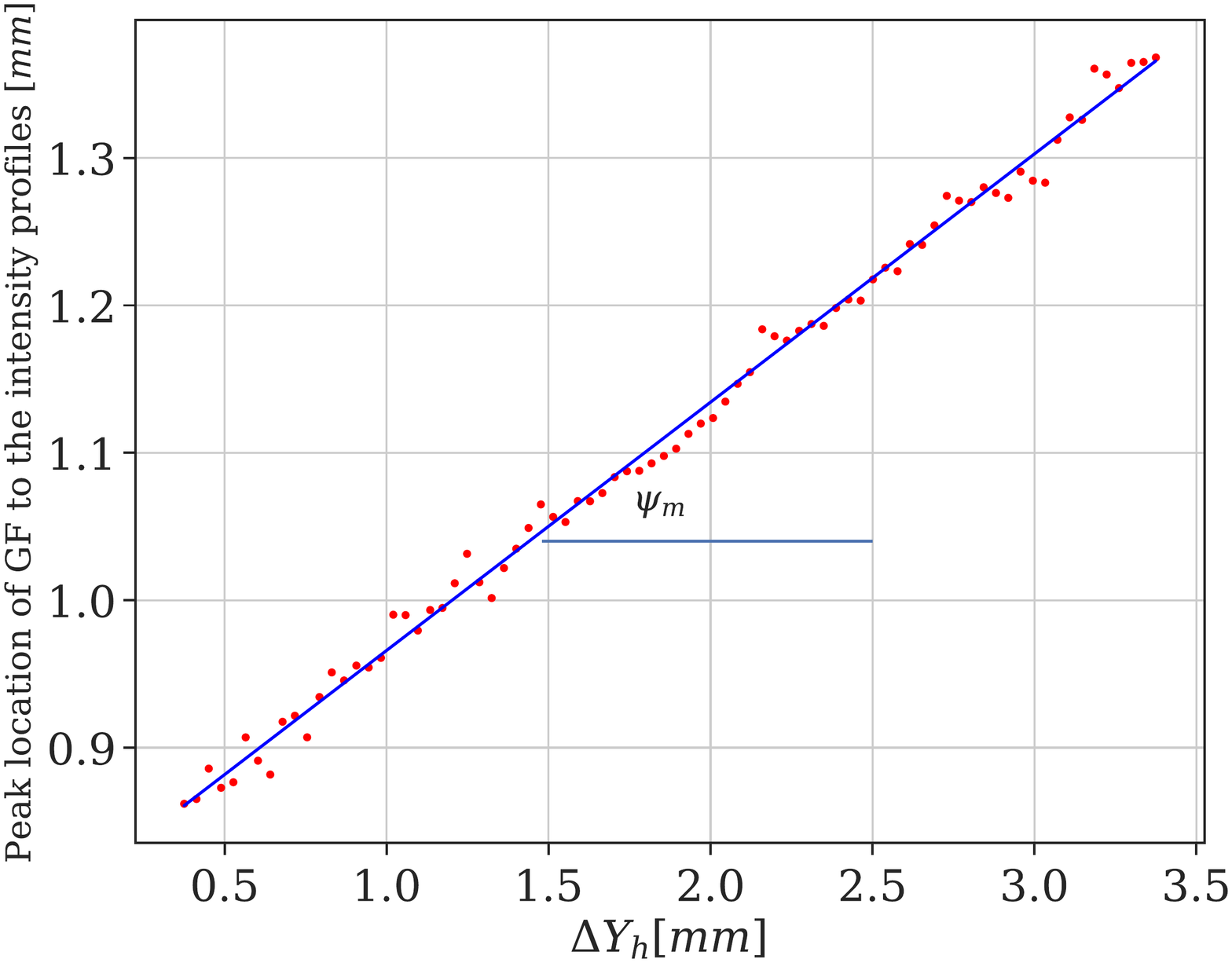}	
\caption{Prediction of the given angle of sheet rotation from the peak locations of the Gaussian fit ($g(x) =  \frac{1}{\sigma \sqrt{2\pi}} e^{-\frac{1}{2} (\frac{x-\mu}{\sigma} )^2}$) curves to the intensity profiles at multiple locations. The blue line is the line fit to the peak locations.}  
\label{fig:peak_locations}
\end{figure}

\begin{figure}
  \centering
      \includegraphics[trim={0cm 0cm 0cm 0cm},clip,width=0.75\textwidth]{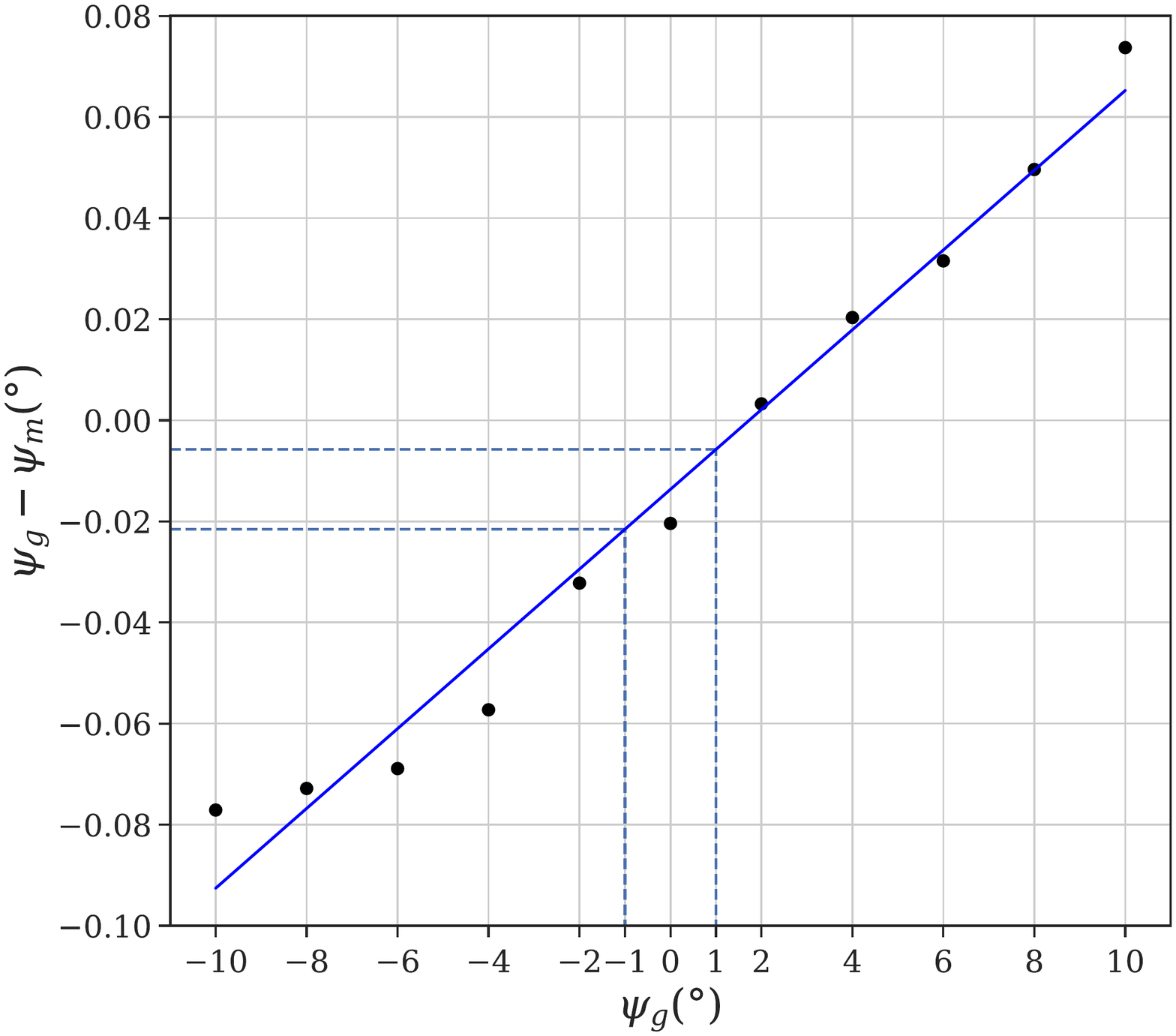}
     \caption{Error in the measured angle  ($\Psi_m$ - $\Psi_g$) plotted against the given angle $\Psi_g$}
\label{fig:angle_given-measured}    
\end{figure}

The alignment achieved using PDs is verified by recording images of the laser sheet shining on a bare (with no lens mounted on), low-cost CMOS array (ArduCAM AR0134 USB3.0 camera shield) which is traversed vertically from $Y_h = +62 \, mm$ to $Y_h = -62 \, mm$ along the front side of the target mount. A traverse is designed for this purpose which slides on the front face of the calibration target mount and can be tightened with a screw to hold it in place as shown in figure \ref{fig:total_alignment_setup}(b). During this verification process, caution must be taken: the irradiance of the laser sheet shining on the camera sensor array should be sufficiently reduced using neutral density filters (NDFs) so that it does not burn the bare camera sensor array. While traversing, each recorded image is convoluted with a Gaussian filter with a $5px$ kernel and a laser sheet profile is plotted across the sheet in the middle of the image. This profile is fitted with a Gaussian curve and the peak location is recorded and later plotted against the vertical distance $\Delta Y_h$ travelled by the camera traverse. A line is fitted through the recorded peak locations, the gradient of which is a measure of the angle of the laser sheet rotation with respect to the reference position. This idea has been tested by rotating an image of the laser sheet recorded at the reference position, from $-10 \degree$ to $+10 \degree$ about $X$ axis and plotting the spatial intensity profiles (SIPs) across the sheet at several locations (some of which are shown by yellow lines show in figure \ref{fig:angle_given_and_SIP_locations}). The peak location of their Gaussian fit curves is plotted against $\Delta Y_h$ as shown in figure \ref{fig:peak_locations}. The gradient of the line fit estimates the given angle with an accuracy of $0.08 \degree$ as shown in figure \ref{fig:angle_given-measured} where the difference of the given and measured angles ($\Psi_g - \Psi_m$) is plotted against the given angle $\Psi_g$. As the maximum allowable laser sheet rotation for the current sheet alignment setup is less than $\pm1 \degree$, the absolute error in the rotation angle prediction in this range is below $0.03 \degree$. This shows the effectiveness of the use of the low-cost array as a method of verification of alignment achieved with the primary method {\em i.e.} by using PDs. 

This process allows us to quantify the alignment at a level where the angle of sheet rotation and the $Z$ offset are less than $0.1 \degree$ and $0.1 \, mm$, respectively. The impact of the uncertainty in these quantities will be quantified in section \ref{sec:validation}. The reference position $\Psi_0$ of the laser sheet was taken as the position of alignment to the calibration target. 

\section{Experimental Validation of the new alignment method in SPIV}
\label{sec:validation}

An experiment has been designed to validate the application of the novel alignment method in 3C-2D SPIV. After the laser sheet is aligned with the calibration target, a suitable mapping function that maps the object space to the image space is computed, the residual errors in the 3C-2D SPIV measurements are quantified and their values are compared with those found in the literature. 

The experiment uses a 3D calibration to dewarp the particle images obtained with two stereo cameras. The particle images are obtained from a transparent box containing micro-particles which are illuminated with the aligned laser sheet and translated by several sets of the known true displacements $T_X,T_Y$ and $T_Z$ in $X,Y$ and $Z$ axes. The 3C displacements $U, V$ and $W$ are reconstructed from the 2C displacements $(u_1, v_1)$ and $(u_2, v_2)$ of camera 1 and camera 2, respectively and are compared with the true displacements $T_X,T_Y$ and $T_Z$ and the residual errors are computed.

\begin{figure}
\begin{center}
\begin{tabular}{c}
\includegraphics[width=\textwidth]{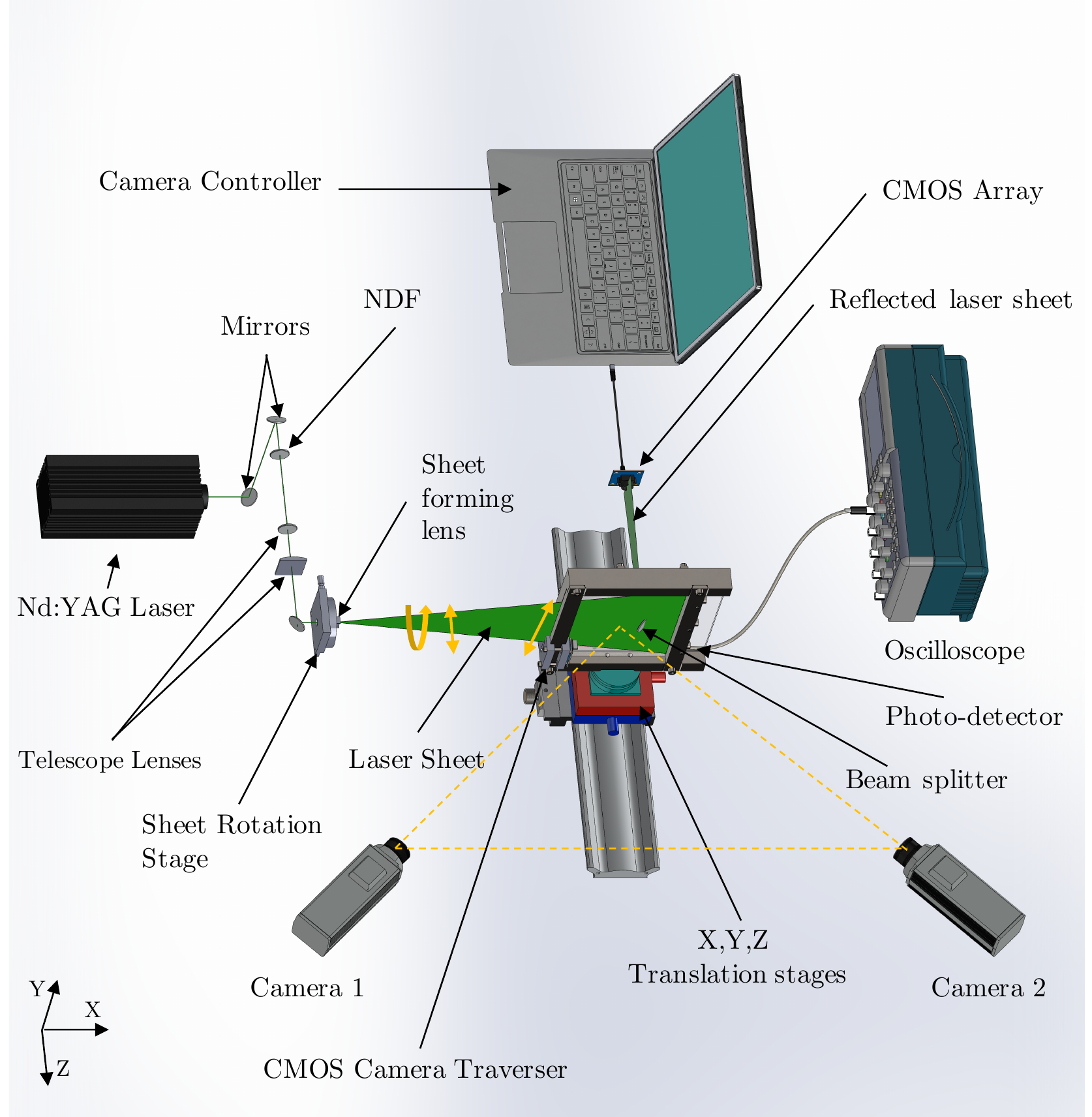} \\
\end{tabular}
\end{center}
\caption{A 3D model of the experimental setup}
\label{fig:experimental_setup}  
\end{figure}

A parametric view of the experimental setup is shown in figure \ref{fig:experimental_setup}. A $2 \, mm$ thick calibration target (made of transparent glass) of about $200 \times 200 \; mm^2$ size, engraved with $1 \pm 0.01 \, mm$ chequerboard pattern on one flat face is fixed in the calibration target mount. The target mount is supported on three translation stages which together are further mounted on an X95 traverse. These stages are used to translate the mount in $X, Y$ and $Z$ axes for fine adjustments to a known position and to apply true known displacements with an accuracy of $100 \mu m$. The X95 traverse is used, if required, for coarse adjustments in the $Z$ axis.

A laser beam of 532 nm wavelength from a New Wave Nd:YAG laser is reduced to a 1 $mm$ diameter by optics following the telescope principle. The beam is then directed towards the FOV (inside the target mount) using three high energy mirrors and then converted to a laser sheet by a concave cylindrical lens of $-9.7mm$ focal length, all mounted on an optomechanical cage system. 

A resin cast ($181.0 \times 141.0 \times 16.5 \; mm^3$) with 1:1 ratio of the resin and the hardener in which Vestosint particles of size $\approx 54\mu m$ are mixed during casting, is used as the particle box. The thickness of the resin from the front face of the particle box to the laser sheet illuminating particles inside it is $8.0mm$. Another cast ($181.0 \times 141.0 \times 8.0 \; mm^3$) is made with the same material but without particles and is put in front of the aligned calibration target while recording the calibration images. This is to account for the change of the refractive index when imaging the illuminated particles inside the transparent resin cast.

\subsection{Calibration}
\label{subsec:calibration}

The alignment between the laser sheet and the calibration target is achieved following the procedure described in section \ref{sec:new_alignment_method}. Then, the calibration target mount is translated by $Z=+8.5 \pm 0.01 mm $ so that the target is located exactly at the middle of the laser sheet. The images of the calibration target are then recorded with two PCO pixelfly cameras, each mounted on a Scheimpflug adapter along with a 55 $mm$ focal length Micro-NIKKOR lens to achieve stereo imaging. The optical axes of the cameras are at $+45\degree$ and $-45\degree$ to the plane of the calibration target. The adapter allows adjustment of a camera and its lens independent of each other and thus enables it to satisfy the Scheimpflug condition \cite{carpentier1901improvements}. Each camera has a $1280 \times 1024 px$ sensor and a pixel size of $6.45 \mu m$. The $f=55 \, mm$ lenses were operated with an f-stop of $2.8$.

The images of the calibration target at $Z =[-0.5, 0.0,0.5]\pm0.01 \; mm$ are recorded with the two stereo cameras. These images are then masked with zero intensity values outside a four-sided polygon to neglect areas outside the region of interest (ROI). Detection of the chequerboard corners is performed using the following method: 

\setlength{\leftskip}{1cm} Edges of the chequerboard pattern in the calibration image are found by applying the Sobel filter  \cite{Sobel2014Operator}. This image of the edges is then cross-correlated with a square marker template (a `$+$' marker with the thickness equal to one pixel and size approximately equal to the size of each chequerboard box). The cross-correlated image has peak intensities at the location of the chequerboard corners. Applying a Gaussian filter with a kernel of the standard deviation of 1 px converts the pixel intensity distribution at peaks to Gaussian and a dot detection algorithm identifies the centre location of this distribution with sub-pixel accuracy. The uncertainty in corner detection in the calibration images is below $0.19 px$. This uncertainty originates from the inaccurate positioning of the chequerboard on the calibration target or the incorrect detection of locations of the chequerboard corners on the calibration targets. Also, the error in identifying the location of these markers is proportional to the size of the markers \cite{fouras2008target}. 

\setlength{\leftskip}{0cm}
The coordinates of the detected corners are sorted such that the dot near the origin of the image is the first (image point) in the sorted list. Placing the sorted image points from each of the three calibration images in the following order makes a 2D image space. 

\begin{equation}
P_{I} \\
=\begin{bmatrix} 
x(Z=-0.5) & y(Z=-0.5)\\
x(Z=0.0) & y(Z=0.0)\\
x(Z=0.5) & y(Z=0.5)\\
\end{bmatrix}[px]
\label{eq:reconstruction_1}
\end{equation}

A 3D object space is created as follows.
\begin{equation}
P_{O} \\
=
\begin{bmatrix} 
X(Z=-0.5) & Y(Z=-0.5) & -0.5 \\
X(Z=0.0) & Y(Z=0.0) & 0.0 \\
X(Z=0.5) & Y(Z=0.5) & 0.5 \\
\end{bmatrix}[mm] \times M [px/mm]
\label{eq:reconstruction_2}
\end{equation}
where $M$ is the magnification factor. To map the 3D object space to the 2D image space, the Soloff method (a least-square polynomial with cubic dependencies in $X$, and $Y$ and quadratic dependency in $Z$) has been used. The advantage of this model over other 3D calibration models (e.g. the pinhole model) is that the imaging parameters such as the magnification factor, focal length etc. need not be determined. This polynomial function also accounts for the lens distortions or other image non-linearities \cite{raffel2018particle}. making it more suitable than the pinhole camera model for calibration in case of lens distortions \cite{wieneke2005stereo}. 

\begin{figure}[tbph]
\begin{center}
\begin{tabular}{cc}
(a) & (b) \\
\includegraphics[width=0.5\textwidth]{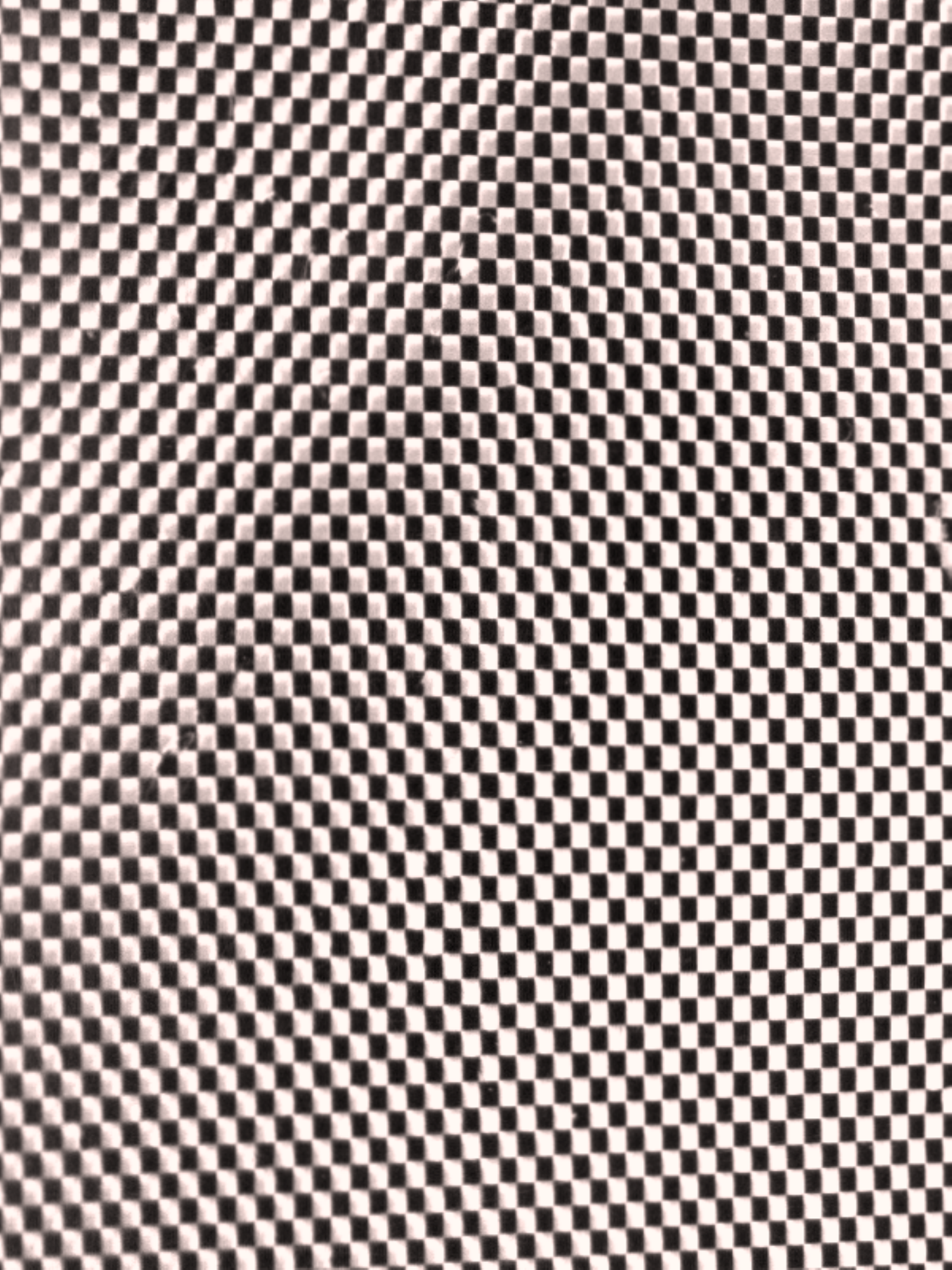} &	\includegraphics[width=0.5\textwidth]{Fig12.pdf}	\\
\end{tabular}
\end{center}
\label{fig:dewarped_calibration_image}	
\caption{ (a) Raw calibration image recorded with camera 1 (b) Dewarped image of (a)}  
\label{fig:raw_and_dewarped_calibration_image}
\end{figure}

The third-order polynomial function in $X$ and $Y$ (referred to as P332 here) is given as follows. 
\begin{equation}
\begin{aligned}
x_i = a_{01} + a_{02} X_i + a_{03} Y_i +  a_{04} Z_i + a_{05} X_i^2 + a_{06} X_i Y_i  + a_{07} Y_i^2 + a_{08} X_i Z_i  + a_{09} Y_i Z_i  + a_{10} Z_i^3 + a_{11} X_i^3 + \\a_{12} X_i^2 Y_i + a_{13} X_i Y_i^2 + a_{14} Y_i^3 + a_{15} X_i^2 Z_i + a_{16} X_i Y_i Z_i + a_{17} Y_i^2 Z_i  + a_{18} X_i Z_i^2  + a_{19} Y_i Z_i^2  , \\
y_i = a_{21} + a_{22} X_i + a_{23} Y_i +  a_{24} Z_i + a_{25} X_i^2 + a_{26} X_i Y_i  + a_{27} Y_i^2 + a_{28} X_i Z_i  + a_{29} Y_i Z_i  + a_{30} Z_i^3 + a_{31} X_i^3 + \\a_{32} X_i^2 Y_i + a_{33} X_i Y_i^2 + a_{34} Y_i^3 + a_{35} X_i^2 Z_i + a_{36} X_i Y_i Z_i + a_{37} Y_i^2 Z_i  + a_{38} X_i Z_i^2  + a_{39} Y_i Z_i^2  \end{aligned}
\label{eq:P332}
\end{equation}

where $(X_i, Y_i, Z_i)$ and $(x_i, y_i)$ are coordinates of the $i^{th}$ point in the object space and image space, respectively. The dewarped image is obtained by interpolating the raw image intensities using the computed mapping functions P332. Figure \ref{fig:raw_and_dewarped_calibration_image} shows the raw and dewarped calibration images. The magnification factor for interpolation is chosen trivially in a manner that the pixel size in the dewarped image is equal to or smaller than that in the raw image to avoid information loss by pixel binning \cite{wieneke2005stereo}.

In the above example, $48\times46 \times 3 = 6,624$ image points along with corresponding object points have been used to compute the mapping function P332. The chosen value of the $M$ is $15\,px/mm$. The mean error ($E$) and the RMS error $\sigma$ are calculated using the following equations:
\begin{equation}
E_x =  \frac{1}{N} \sum_{i=1}^{N}  (\hat{x_i} - x_i) 
\end{equation}

\begin{equation}
\sigma_x =  \sqrt{\frac{1}{N} \sum_{i=1}^{N} ( \hat{x_i} - x_i - E_x)^2}
\end{equation}

where $\hat{x}_i$ is the $x$ coordinate of the point mapped from the $i^{th}$ object point to the corresponding image point and $N$ is the total number of points.  

The values of the maximum absolute, the mean and the RMS error in the nonlinear least-squares fit using P332 are given in table \ref{tab:fit_error}. The maximum absolute error and the RMS error are less than a pixel, which shows that the mapping has a sub-pixel accuracy.

\begin{table}
\caption{The mean and RMS error in the non-linear least squares fit of P332}
\label{tab:fit_error}      
\centering
\begin{tabular}{ccc}
\hline\noalign{\smallskip}
 &  \textbf{in $\bm{x} [px]$ }  &\textbf{ in $\bm{y} [px]$ }   \\
\noalign{\smallskip}\hline\noalign{\smallskip}
Maximum error       & 0.1769        &0.2305 \\ 
$E$     			&0.0002     	&-0.0001  \\ 
$\sigma$        	&0.0439 		&0.0373   \\ 
\noalign{\smallskip}\hline
\end{tabular}
\end{table}

\begin{figure}
  \centering
      \includegraphics[width=\textwidth]{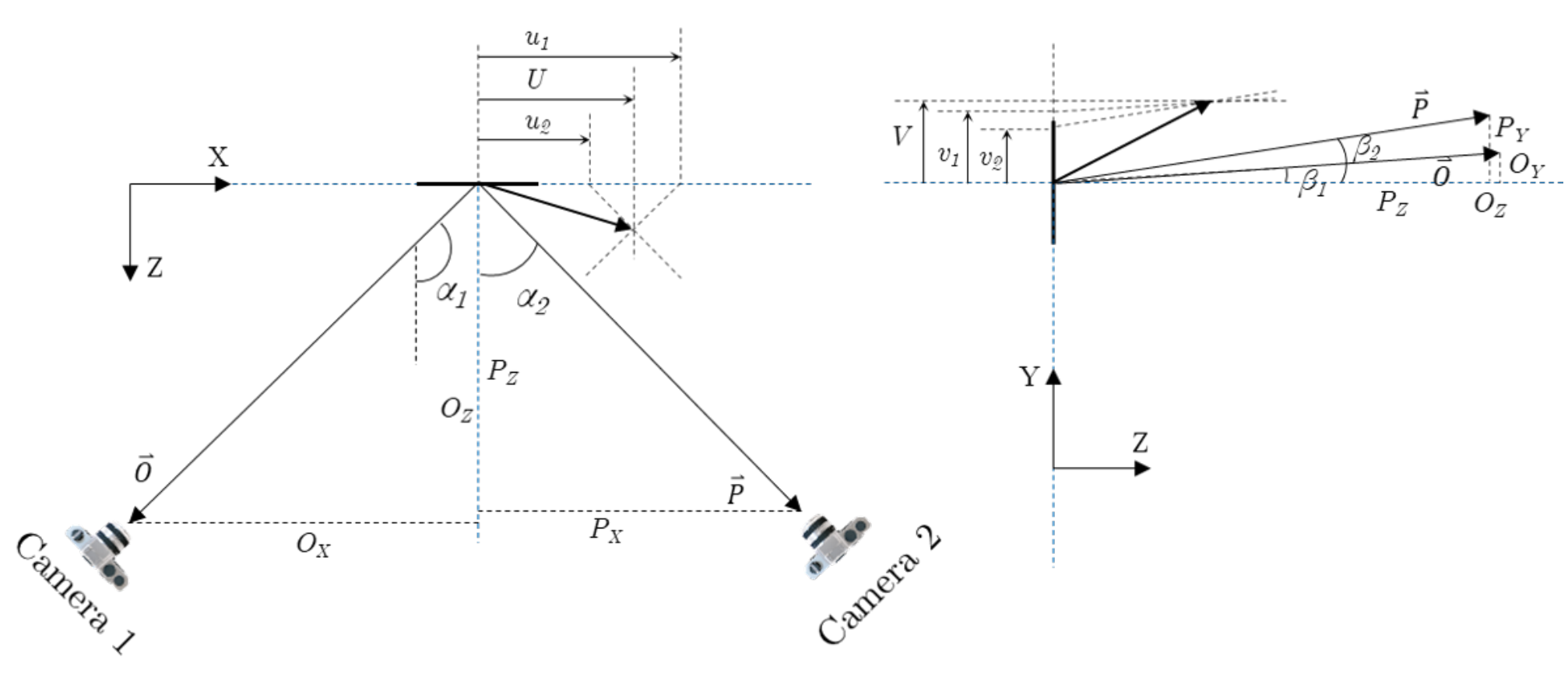}
     \caption{A schematic of 3C reconstruction from 2C displacement vector, derived from \cite{willert1997stereoscopic} and \cite{giordano2006correction}.} 
\label{fig:3C_reconstruction}    
\end{figure}

The particle box is translated in $X,Y$ and $Z$ axes (within the thickness of the light sheet) to apply true displacements using the three translation stages. The step-change in position is chosen to be $500\, \mu m$ in the $X$ and $Z$ axes and $380\, \mu m$ in $Y$ axis. Each set of the true displacements $(T_X, T_Y, T_Z)$ is given in table \ref{tab:final_results}. The images are then dewarped with P332 computed in section \ref{subsec:calibration}. The dewarped particle images are then individually paired with the one at $T_X,  T_Y, T_Z=[0,0,0]$. Multigrid/multipass 2C-2D digital PIV cross-correlation \cite{soria1996investigation} of the image pairs is performed to compute the $(u_j,v_j)$ displacement vectors with respect to $X,Y,Z=(0,0,0)$ where $j=\{1,2\}$ for the two cameras. The 3C reconstruction is performed to find the particle displacements: $U,V$ and $W$ (in the $X,Y$ and $Z$ axes). Although the 3C-reconstruction technique introduced by \citet{soloff1997distortion}, which involves the gradients of the mapping functions, gives sufficiently accurate results which are comparable to those found in literature as reported in \citet{giordano2009spatial}, the technique of \citet{willert1997stereoscopic} was used in this paper. The latter technique requires at least three equations to solve the system for the three unknowns. PIV analysis of the dewarped particle images from camera 1 and camera 2 gives four displacement components, $u_1$, $v_1$ and $u_2$, $u_2$ respectively. Calculation of the three displacement components $U$, $V$ and $W$ from the four displacements values makes the system overdetermined. This technique also uses the angles $\alpha_1$, $\alpha_2$, $\beta_1$ and $\beta_2$ as shown in figure \ref{fig:3C_reconstruction} for the 3C reconstruction. The equations to find the three components $U,V$ and $W$ are as follows:

\begin{equation}
U = \frac{u_1 \tan \alpha_2 - u_2 \tan \alpha_1}{\tan \alpha_2 - \tan \alpha_1}
\end{equation}
\begin{equation}
V = \frac{v_1 + v_2 }{2} + \frac{u_2 + u_1}{2} \left( \frac{\tan \beta_2  - \tan \beta_1}{\tan \alpha_1  - \tan \alpha_2 } \right)
\end{equation}
\begin{equation}
W = \frac{u_2  - u_1 }{\tan \alpha_1 - \tan \alpha_2} = \frac{v_2  - v_1 }{\tan \beta_1 - \tan \beta_2}
\end{equation}

As $\beta_1$ and $\beta_2$ are small for small $Y$ displacements, the tangents of these values are also small and can be considered as zero without introducing any considerable errors in the measured displacements. The accurate measurement of $\alpha_1$ and $\alpha_2$ from the geometry of the experimental setup is hard to achieve. \citet{fei2004investigations} suggested the following formulation to measure the local angles $\alpha_1$ and $\alpha_2$.

\begin{equation}
\tan \alpha = \frac{dZ ^\prime _m}{dZ} \cdot \frac{dX}{dX ^ \prime _m}
\end{equation}

where ${dZ^\prime}_m = M \cdot dZ^\prime$,   $dZ ^\prime  = dZ \cdot \tan \alpha $ is the projection of a known displacement $dZ$ onto the measurement plane, $dX^\prime$ is of $dX$ and $M$ is the magnification factor. 

As the 3C-2D measurements were obtained after aligning the laser sheet with the calibration target, no ad hoc post-correction scheme is applied to these measurements.

\subsection{Results and Discussion}
\label{subsec:results_and_discusion}

\begin{figure}[tbph]
\begin{center}
\begin{tabular}{c}
\includegraphics[width=\textwidth]{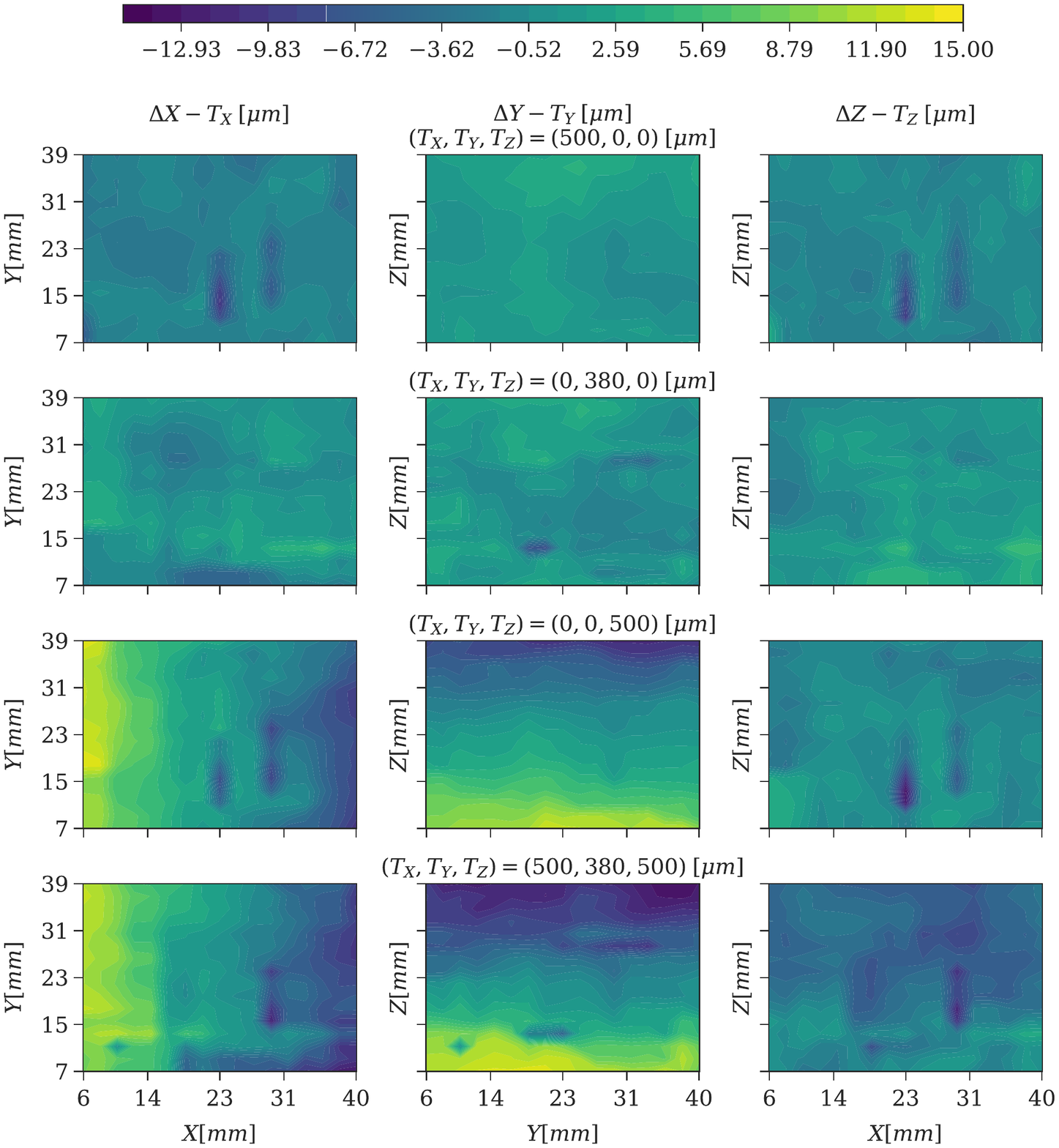} \\ 
\end{tabular}
\end{center}
\vspace*{-0.3in}\caption{
(a) Contour maps of the absolute error in displacements after the manual translations of the particle box presented at serial number 5, 2, 1, and 9 in table \ref{tab:final_results}.}
\label{fig:final_results_contours}
\end{figure}

The mean and the RMS errors in the 3C-2D SPIV measurements after the manual translations of particle box are given in table \ref{tab:final_results}. The absolute error values are below $14\mu m$ which is comparable to those found in literature {\em e.g.} in  \citet{soloff1997distortion}. The largest absolute error in these measurements is $1/3200$ the size of the mapping function domain which is the size of the calibration target plane used to compute the mapping function. The absolute error is larger for the out-of-plane displacements than the in-plane displacements. Figure \ref{fig:final_results_contours} shows the contour maps of the absolute errors in the SPIV measurements after the manual translations presented at the number 5, 2, 1, and 9 in table \ref{tab:final_results}. The bias error in measurements 1 and 9 is due to the out-of-plane displacements for which half of the laser sheet thickness is outside the Z range of the 3D calibration. Therefore, the calibration must span the out-of-plane displacement range. Interestingly, the bias error appears in the $X$ and $Y$ axes but not in $Z$ axis for the true out-of-plane displacements. The bias error was not found in the in-plane measurements.

The residual error in the above measurements arises from the following error sources:

\begin{enumerate}
\item Uncertainty in detecting the exact locations of chequerboard corner points (i.e. image points). 
\item Calibration errors in the computation of mapping functions. 
\item Reconstruction error, defined as the residual of the least-squares fit in 3C vector reconstruction from 2C displacement vectors. 
\item Instrument resolution: the resolution of the translation stages used to apply the true displacements 
\item The parallax error in reading the micrometre gauge while applying the true displacements. 
\end{enumerate}

A combination of these error sources leads to the maximum uncertainty of $7.6 \, \mu m$ (equivalent to $0.114\, px$) in the SPIV measurements in this experiment which is comparable to the uncertainty of $0.11 \, px$ achieved with the self-calibration method \cite{wieneke2005stereo}. It is also below the resolution of micrometre translation stages. This is also below the uncertainty in the detection of the chequerboard corners in the calibration images and the maximum error in mapping the object space to the image space. This reinforces the point that there is no need for any ad hoc post-correction scheme to the 3C-2D SPIV displacements to account for the misalignment errors if the SPIV measurements are taken after aligning the laser sheet with the calibration target using the new alignment method.

\newpage

\begin{table}
\caption{The true displacements, and the absolute and RMS errors in measured displacements for particle images dewarped with P332 mapping function}
\label{tab:final_results}
\centering
\begin{tabular}{ccccccc}
\hline\noalign{\smallskip}
No. & $T_X\pm10(\mu m)$   & $T_Y\pm10(\mu m)$  & $T_Z\pm10(\mu m)$   &$E_X(\mu m)$ &$E_Y(\mu m)$ &$E_Z(\mu m)$ \\ 
&           &       &         &$\sigma_X(\mu m)$ &$\sigma_Y(\mu m)$ & $\sigma_Z(\mu m)$ \\
\noalign{\smallskip}\hline \hline\noalign{\smallskip}
1 & 0  & 0 & 500       & 1.3   & 0.8  & -0.1  \\ 
&           &       &       & 6.0  & 6.0  & 2.0  \\ \hline
2 & 0  & 380 & 0       & 0.1  & 0.5  & 0.7  \\ 
&           &       &       & 0.8   & 0.7   & 0.7   \\ \hline
3 & 0  & 380 & 500       &  2.3  & -1.5   & 0.4   \\ 
&           &       &       & 6.6  & 7.5  & 2.8   \\ \hline
4 & 0  & 380 & -500      & -0.4   & 2.2   & 1.0  \\ 
&           &       &       & 6.4  & 7.0   & 1.8  \\ \hline
5 & 500  & 0 & 0       & -2.2   & 1.1  & -1.3  \\ 
&           &       &       & 1.4  & 1.1  & 1.5   \\ \hline
6 & 500  & 0 & 500       & 2.2  & -1.2  & -0.5    \\ 
&           &       &       & 7.1  & 5.7 &  1.7   \\ \hline
7 & 500  & 380 & 0       & 0.2  & -1.2  & -2.0    \\ 
&           &       &       & 2.4  & 1.6  & 2.4    \\ \hline
8 & 500 & 380 & 500        &0.5  & -1.2  & -3.7     \\ 
&           &       &       & 6.5  & 7.6  & 2.6    \\ \hline
9 & 500 & 380 & -500       & -2.4  & -1.4  &  1.4     \\ 
&           &       &       & 6.4  & 7.1 &  1.9     \\ \hline
\noalign{\smallskip}\hline
\end{tabular}
\end{table}

\newpage
\section{Summary}
\label{sec:summary}
A precise alignment of the laser sheet with the calibration target is required in commonly used methods of calibration for 2C-2D planar PIV and 3C-2D SPIV measurements. The misalignment causes \textit{perspective error} in 2C-2D planar PIV and \textit{position error} and \textit{3C-reconstruction error} in 3C-2D SPIV. While the precise alignment is difficult to achieve, various methods including self-calibration and target-free calibration have previously been proposed to correct or avoid the errors arising mainly from this misalignment. These methods are prone to create bias in the overall results because the correction schemes mostly require a disparity map which is computed from the cross-correlation of particle images and can be erroneous due to the loss of point-correspondence between the images. The loss of point-correspondence also leads to erroneous computation of the mapping functions in the target free calibration. Hence, a more reliable method is required to avoid misalignment errors. The method proposed in this paper focuses on the basic requirement, the \textit{alignment} between the laser sheet and calibration target. This technique uses a locally fabricated calibration target mount, to hold the calibration target in place so that the laser sheet could be steered to be aligned to the target mount. The alignment procedure includes measuring the intensity of the laser sheet with three photo-detectors fixed in the same target mount at three vertically equidistant positions. The accuracy of this alignment procedure is verified by imaging the laser sheet with a low-cost CMOS camera, traversed along the front face of the target mount. The recorded images of the aligned calibration target are used to compute a polynomial mapping function introduced by \citet{soloff1997distortion} with an uncertainty of less than $0.1\, px$, which is then used to dewarp the particle images. This function is also able to automatically correct the geometric distortions in the particle images. 

An experiment has been performed with a particle box that is given known true displacements in the object space. The displacements measured at the image planes of two cameras, following a 3C reconstruction, show the uncertainty of less than $07.6 \, \mu m$ (or $0.114\, px$) which is below the uncertainty in detecting the chequerboard corners detection, the maximum error in mapping the object points to the image points and the uncertainty in the applying the true displacements. This is also is comparable to the uncertainty of $0.11 \, px$ of the self-calibration method as reported by \citet{wieneke2005stereo}. The largest absolute error is $1/3200$ the size of the mapping function domain. This is achieved without any ad hoc post-correction to the SPIV measurements. The results show the effectiveness and applicability of the new method of alignment to minimize the errors associated with misalignment.

\begin{acknowledgements}
The authors would like to acknowledge the support of the Australian Government for this research through an Australian Research Council Discovery grant. C. Atkinson was supported by the ARC Discovery Early Career Researcher Award (DECRA) fellowship. Muhammad Shehzad also acknowledges the Punjab Educational Endowment Fund (PEEF), Punjab, Pakistan for funding his PhD research. Sean Lawrence gratefully acknowledges the financial support of the Maritime Division of the Defence Science and Technology Group. 

\end{acknowledgements}

\newpage

\FloatBarrier

\bibliographystyle{elsarticle-harv} 
\bibliography{Paper-SPIV}

\end{document}